\documentclass[runningheads]{llncs}
\pdfoutput=1
\usepackage{graphicx}
\usepackage[utf8x]{inputenc}
\usepackage{algorithm,algorithmic}
\usepackage{amsmath,amssymb,amsfonts}
\usepackage{textcomp}
\usepackage{xcolor}
\begin{document}
\title{ A New Fairness Model based on User's Objective for Multi-user Multi-processor Online Scheduling}
%
%
\author{Debasis Dwibedy\inst{1}\and
Rakesh Mohanty\inst{1}}
\authorrunning{D. Dwibedy and R. Mohanty}
%
\institute{Veer Surendra Sai University of Technology, Burla 768018, India 
\email{debasis.dwibedy@gmail.com}\\
\email{rakesh.iitmphd@gmail.com}}
\maketitle              
\begin{abstract}
Resources of a multi-user system in multi-processor online scheduling are shared by  competing users in which fairness is a major performance criterion for resource allocation. Fairness ensures equality in resource sharing among the users. According to our knowledge, fairness based on the user's objective has neither been comprehensively studied nor a formal fairness model has been well defined in the literature. This motivates us to explore and define a new model to ensure algorithmic fairness with quantitative performance measures based on optimization of the user's objective. In this paper, we propose a new model for fairness in \textit{Multi-user Multi-processor Online Scheduling Problem(MUMPOSP)}. We introduce and formally define quantitative fairness measures based on user's objective by optimizing makespan for individual user in our proposed fairness model. We also define the unfairness of deprived users and absolute fairness of an algorithm. We obtain lower bound results for the absolute fairness for $m$-identical machines with equal length jobs. We show that our proposed fairness model can serve as a framework for measuring algorithmic fairness by considering various optimality criteria such as flow time and sum of completion times. 
\end{abstract}
\section{Introduction}
\label{sec:Introduction}
\begin{figure}[ht]
\includegraphics[scale=0.7]{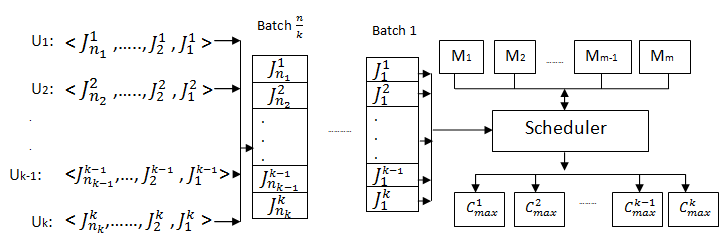}
\caption{Illustratopn of MUMPOSP for k Users with Equal Number of Jobs}
\label{fig:mumposp.png}  
\end{figure}

Supercomputers, grids and clusters in High Performance Computing(HPC) and web servers in client-server networking have gained immense practical significance as Service Oriented System(SOS) in modern day computation [1]. Unlike the traditional computing system such as personal computer, the \textit{SOS} supports multiple users. The users compete for system's resources for execution of their jobs. The most popular cluster scheduler \textit{MAUI} [2] and the well-known \textit{BOINC} platform [3] deal with a number of competing users, where each user submits  a set of jobs simultaneously and seeks for a \textit{minimum time of completion(makespan)} for their respective submissions. A non-trivial challenge for the \textit{SOS} is to ensure fair scheduling of jobs for multiple users while optimizing their respective makespans.\\
In this paper, we define and quantify fairness as a performance measure based on user's objective for any resource allocation algorithm in multi users \textit{HPC} systems. Particularly, we consider processors as the resources to be shared and makespan as the user's objective in the \textit{Multi-user Multi-processor Online Scheduling Problem(MUMPOSP)}. We formally define the \textit{MUMPOSP} as follows.\\\\
\textbf{Multi-user Multi-processor Online Scheduling Problem(MUMPOSP)}
\begin{itemize}
\item \textbf{Inputs:} We are given a set of $m$ identical processors, i.e. $M$=$\{M_1, M_2,..,M_m\}$ and a set of $n$ jobs, where $m\geq 2$ and $n>>>m$. Let $U_r$ represents a \textit{user}, where $1\leq r\leq k$ and $J^{r}$ is the sequence of jobs requested by \textit{user} $U_r$, where $J^{r}$=$(J^{r}_i|1\leq i\leq n_r)$ such that $J$=$\bigcup_{r=1}^{k}{J^r}$, $\sum_{r=1}^{k}{n_r}$=$n$ and $J^x\cap J^y$=$\phi$, where $x\neq y$ and $1\leq (x, y)\leq k$. The processing time of job $J^{r}_i$ is $p^{r}_i$, where $p^{r}_i\geq 1$.
\item \textbf{Output:} Generation of a Schedule($S$) in which makespan for each user ($U_r$) is denoted by  $C^{r}_{max}$=$\max\{c^{r}_i|1\leq i\leq n_r\}$, where $c^{r}_i$ is the completion time of job $J^{r}_{i}$ 
\item \textbf{Objective:} Minimize $C^{r}_{max}$, \hspace*{0.1cm}  $\forall U_r$.
\item \textbf{Constraint}: The scheduler can receive a batch of at most $r$ jobs at any time step and the jobs must be irrevocably scheduled before the arrival of next batch of jobs, where $1\leq r\leq k$.
\item \textbf{Assumption.} The jobs are independent and are requested from $k$ parallel users, where $k\geq 2$ 
\end{itemize}
\textbf{Illustration of MUMPOSP.}
For simplicity and basic understanding of the readers, we illustrate an instance of \textit{MUMPOSP} for scheduling of $n$ jobs that are submitted by $k$ users($U_r$) as shown in \textit{Figure} \ref{fig:mumposp.png}. Here, $M_1$, $M_2$,...,$M_m$ represent $m$ identical machines and $U_1$, $U_2$,...,$U_{k-1}$, $U_k$ denote job sequences for $k$ users. Here, each user has $\frac{n}{k}$ jobs. Jobs are submitted in batches online, where a batch is constructed after receiving exactly one job from each user(as long as an user has an unscheduled job). A batch consists of at least one job. Therefore, we have at least $1$ batch, where $k$=$n$ and at most $n-k+1$ batches, where any one of the users $U_r$ has $n_r$=$n-k+1$ and remaining users have exactly one job each. Each user($U_r$) seeks to obtain a minimum value for its makespan($C^{r}_{max}$) as the output, rather than the overall makespan($C_{max}$) of the system. Hence, it is indispensable for the scheduler to be fair while optimizing the $C^{r}_{max}$ for each user.\\
\textbf{Representation of MUMPOSP.} By following general framework $\alpha|\beta|\gamma$ of Graham et al.[16], we represent \textit{MUMPOSP} as $MUMPOSP(k, P_m|C^{r}_{max})$, where $P_m$ denotes $m$-identical machines and $k$ is the number of users.\\\\
\textbf{Perspectives of Fairness.}
Fairness has been considered as a major performance criterion for the scheduling algorithms in multi-user systems [4, 5]. Fairness has been studied from two perspectives such as allocation of resources to the users and user's objective. Fairness of an algorithm with respect to resource allocation, guarantees uniform allocation of resources to the competing users [6]. The resources to be shared are application dependent. For example, in client-server networking, the shared resources may be throughput or network delay or specific time quantum [7, 8], whereas in distributed systems, the resources may be processors or memory or time slice [9, 10]. \\
Algorithmic fairness based on user's objective is evaluated by the objective values achieved for respective users. An equality in the obtained objective values for a user ensures fairness of a scheduling algorithm. It is important for a fairness measure to define the equality for quantifying how far an achieved objective value is far from the defined equality.\\\\
\textbf{Related Work.} Fairness as a quantitative performance measure based on resource allocation was comprehensively studied by Jain et al. [6]. A set of properties for an ideal fairness measure was defined and a fairness index $F(x)$ was proposed for resource allocation schemes. If any scheduling algorithm assigns resources to $k$ competing users such that $r^{th}$ user gets an allocation of $x_r$. then $F(x)$ is defined as follows.\\\\
\hspace*{3.2cm}$F(x)$=$\frac{[{\sum_{r=1}^{k}{x_r}}]^2}{\sum_{r=1}^{k}{{x_r}^2}}$,\hspace*{1.2cm} where $x_r\geq 0$. \\\\
The value of $F(x)$ is bounded between $0$ and $1$ to meaningfully show percentage of fairness and discrimination of a resource allocation scheme for each user. Fairness based on sharing of resources such as processors, memory, system clock and system bus in multi-programmed multi-user system was well studied in [9-11]. Some recent works with contributions on algorithmic fairness of online scheduling can be found in [17-18]. According to our knowledge, study of fairness based on user's objective has not been exhaustively studied in the literature.  \\
In [12-15], \textit{stretch} matrix has been considered as a user's objective based fairness measure for resource scheduling algorithms in multi-user system. Stretch($d^{r}_{A}$) is defined as a degradation factor in the objective value obtained by any algorithm \textit{A} for each user $U_r$. Let us consider $V^{r}_{A}$ is the objective value achieved by algorithm \textit{A} and $V^{r}_{OPT}$ is the optimum objective value for respective $U_r$. Now, stretch can be defined as\\
\hspace*{3.2cm} $d^{r}_{A}$=$\frac{V^{r}_{A}}{V^{r}_{OPT}}$\\\\
The objective of any scheduling algorithm is to incur an equal stretch for each $U_r$ to ensure fairness. Stretch matrix guarantees fairness based on equality in achieved objective values. However, it fails to bound the fairness and unable to show the exact value of fairness per user as well as overall fairness of a scheduling algorithm. Further, the discrimination of any scheduling algorithm for the deprived users can not be captured by the stretch matrix. Therefore, it is quintessential to define a formal fairness measure based on user's objective.\\\\      
\textbf{Our Contribution.} In our work, we propose a new model for fairness in \textit{Multi-user Multi-processor Online Scheduling Problem(MUMPOSP)}. We introduce and formally define quantitative fairness measures based on user's objective by optimizing makespan for individual user in our proposed fairness model. We also define the unfairness of deprived users and absolute fairness of an algorithm. We obtain lower bound results for the absolute fairness for $m$-identical machines with equal length jobs. We show that our proposed fairness model can serve as a framework for measuring algorithmic fairness by considering various optimality criteria such as flow time and sum of completion times. 


\section{Our Proposed Fairness Model}
\begin{figure}[ht]
\includegraphics[scale=0.7]{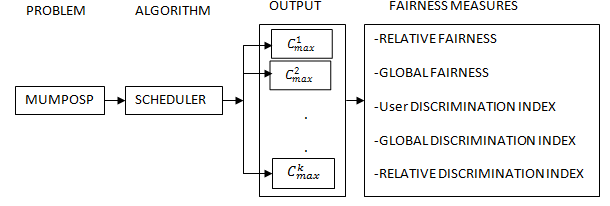}
\caption{A Fairness Model based on User's Objective}
\label{fig:fairnessmodel3.png}  
\end{figure}

We develop a new fairness model in which, we define five quantitative measures to ensure algorithmic fairness. Instead of considering the resource allocation at the input level, the model considers the achieved value of user's makespan at the output level to determine the fairness of a scheduling algorithm. The model captures the issues of relative and global parameters for fairness by a \textit{Fairness Index(FI)}. The issues of unfairness is captured by a \textit{Discrimination Index(DI)}. The \textit{FI} includes fairness measures such as \textit{Relative Fairness(RF)} and \textit{Global Fairness(GF)}. Higher value of any fairness measure indicates more fair algorithm.  The \textit{DI} includes unfairness measures such as \textit{User Discrimination Index(UDI)}, \textit{Global Discrimination Index(GDI)} and \textit{Relative Discrimination Index(RDI)}. Lower value of any unfairness measure indicates more fair algorithm. Before defining the fairness measures, we illustrate our proposed fairness model and discuss the characteristics of a good fairness model as follows.\\\\
\textbf{Illustration of our Proposed Fairness Model.} We illustrate our proposed fairness model as shown in Figure \ref{fig:fairnessmodel3.png}. The model considers the \textit{MUMPOSP} and captures the fairness of online scheduling algorithm by considering the makespan($C^{r}_{max}$) of individual user at the output. \\\\\\
\\\\
\subsection{Characteristics of a Good Fairness Model}
\label{subsec:Characteristics of a Good Fairness Model }
Motivated by the seminal work of Jain et al. [6] for  characterization of a good fairness model, we capture the following essential properties in our model to develop our fairness measures. 
\begin{itemize}
\item \textit{Independent of Input Size}. A  model must be applicable to any number of users irrespective of the number of jobs offered by the users and availability of any number of machines.
\item \textit{Independent of Scale}. The model should be independent of unit or scale of measurement. The model should be able to measure fairness irrespective of the fact that processing time of the jobs are given in seconds or micro-seconds or nano-seconds. The measuring unit must be uniform or inter convertible.  
\item \textit{Finitely Bounded}. The model must bound the value of fairness measure within a finite range, preferably between $0$ and $1$ such that  percentage of fairness for respective users can easily be determined.
\item \textit{Consistent}. If any change in the scheduling policy results in different makespan for at least one user, then the change in the fairness measure must be reflected to the concerned users as well as to the overall fairness of the policy.
\end{itemize} 
In addition to the above mentioned properties from the literature, we also consider relative and overall fairness as en essential feature to develop our fairness measures. The model must represent \textit{relative equality} among achieved objective values for the users to show user's fairness of an algorithm. For example, the users may not seek equal makespan as a gesture of fairness, however, they require to obtain an equal ratio between the \textit{desired makespan(optimum value)} to the achieved makespan  for all users. The value obtained by an algorithm for relative equality leads to \textit{relative fairness} with respect to each user. Also, the model should show \textit{overall fairness} of the algorithm with respect to all users.
  
 
\subsection{Our Proposed Fairness Measures}
By considering the above mentioned desirable properties, we now define formal measures of fairness and unfairness for \textit{MUMPOSP} as follows.\\
If any algorithm \textit{A}, schedules jobs of $k$ competing users on $m$ parallel machines such that $r^{th}$ user obtains a makespan of $C^{r}_{A}$, then we define the following measures. 
\begin{definition}
The \textbf{Relative Fairness(RF)} obtained by algorithm \textit{A} for any user $U_r$ is defined by \\ 
\hspace*{4.0cm} ${RF}(C^{r}_{A})$=$\frac{C^{r}_{OPT}}{C^{r}_{A}}$, \hspace*{1.0cm}where, $C^{r}_{OPT}$=$\frac{\sum_{i=1}^{n_r}{p^{r}_{i}}}{m}$  \hspace*{0.6cm}(1) 
\end{definition}
\begin{corollary}
The \textbf{Relative Fairness Percentage(RFP)} for any user $U_r$ obtained by algorithm \textit{A} is defined by \\
\hspace*{4.0cm} ${RFP}(C^{r}_{A})$=$R(C^{r}_{A})\cdot 100$\hspace*{3.9cm}(2)
\end{corollary}
\begin{definition}
The \textbf{Global Fairness(GF)} of algorithm \textit{A} for $k$ users is defined by\\
\hspace*{4.0cm} ${GF}(C^{r}_{A}, k)$=$\frac{1}{k}\cdot \sum_{r=1}^{k}{(RF(C^{r}_{A}))}$ \hspace*{2.5cm}(3)
\end{definition}
\begin{corollary}
The \textbf{Global Fairness Percentage(GFP)} of any algorithm \textit{A} for $k$ users is defined by \\
\hspace*{4.0cm} ${GFP}(C^{r}_{A}, k)$=$GF(C^{r}_{A}, k)\cdot 100$\hspace*{2.9cm}(4)
\end{corollary}
Again, if any algorithm \textit{A}, schedules jobs of $k$ competing users such that $r^{th}$ user obtains a makespan of $C^{r}_{A}$, then we define \textbf{Fairness Index} for algorithm \textit{A} represented by $2$-tuple with  two parameters such as $RF$ and $GF$ as follows\\\\
\hspace*{1.2cm} $FI(C_A)$=$<\{RF(C^{r}_{A})|1\leq r\leq k\}, GF(C^{r}_{A}, k)>$\hspace*{3.1cm}(5) \\\\
\textbf{Example 1:} Let us consider $3$ users $U_1$, $U_2$ and $U_3$ with jobs $U_1$=$\{J^{1}_{1}/1, J^{1}_{2}/2 \}$, $U_2$=$\{J^{2}_{1}/3, J^{2}_{2}/4 \}$ and $U_3$=$\{J^{3}_{1}/5, J^{3}_{2}/6 \}$ respectively. Suppose any algorithm \textit{A} schedules the jobs of $U_1$, $U_2$ and $U_3$ such that $C^{1}_{A}$=$11$, $C^{2}_{A}$=$9$ and $C^{3}_{A}$=$10$, then we have, ${RF}(C^{1}_{A})$=$\frac{1.5}{11}$=$0.13$ and ${RFP}(C^{1}_{A})$=$13\%$, ${RF}(C^{2}_{A})$=$\frac{3.5}{9}$=$0.38$ and  ${RFP}(C^{2}_{A})$=$38\%$, ${RF}(C^{3}_{A})$=$\frac{5.5}{10}$=$0.55$ and ${RFP}(C^{3}_{A})$=$55\% $. Therefore, we have ${GF}(C^{r}_{A}, 3)$=$0.35$ and ${GFP}(C^{r}_{A}, 3)$=$35\%$. \\
\begin{definition}
The \textbf{Unfairness} of any algorithm \textit{A} for \textit{MUMPOSP} with respect to each user $U_r$ is defined by \textbf{User Discrimination Index} as\\
\hspace*{1.4cm} ${UDI}_{A}^{r}$=$1- {RF}(C^{r}_{A})$\hspace*{7.1cm}(6)
\end{definition}
\begin{definition}
The \textbf{Overall Unfairness} of algorithm \textit{A} for $k$ users is defined by \textbf{Global Discrimination Index} as \\
\hspace*{1.4cm} ${GDI}(C^{r}_{A}, k)$=$1-{GF}(C^{r}_{A}, k)$ \hspace*{5.6cm}(7)
\end{definition}
\begin{definition}
The \textbf{Realtive Discrimination Index(RDI)} of any algorithm \textit{A} for \textit{MUMPOSP} with respect to each user $U_r$ is defined as \\
\hspace*{1.4cm}${RDI}_{A}^{r}$=${GF}(C^{r}_{A}, k)- {RF}(C^{r}_{A})$,  if ${RF}(C^{r}_{A})< {GF}(C^{r}_{A}, k)$ \\
\hspace*{1.4cm}${RDI}_{A}^{r}$=$0$,\hspace*{0.6cm} otherwise \hspace*{6.2cm}(8)
\end{definition}
Again, if any algorithm \textit{A}, schedules jobs of $k$ competing users such that $r^{th}$ user obtains a makespan of $C^{r}_{A}$, then we define \textbf{Discrimination Index} for algorithm \textit{A} as $3$-tuple with three parameters such as $UDI$, $GDI$ and $RDI$ as follows. \\\\
\hspace*{1.2cm} ${DI}(C_A)$=$<\{{UDI}^{r}_{A}|1\leq r\leq k\}, {GDI}(C^{r}_A, k),  \{{RDI}_{A}^{r}|1\leq r\leq k\}>$\hspace*{0.3cm}(9)\\\\
\textbf{Example 2:} Let us consider any algorithm \textit{A} results relative fairness for $U_1$, $U_2$, $U_3$ and $U_4$ as $0.6$, $0.6$, $0.6$ and $0.2$ respectively. We now have ${GF}(C^{r}_{A}, 4)$=$0.5$. Therefore,${UDI}_{A}^{1}$=$1-0.6$=$0.4$, ${UDI}_{A}^{2}$=$1-0.6$=$0.4$, ${UDI}_{A}^{3}$=$1-0.6$=$0.4$, ${UDI}_{A}^{4}$=$1-0.2$=$0.8$, ${GDI}(C^{r}_{A}, k)$=$1-0.5$=$0.5$ and ${RDI}_{A}^{4}$=$0.5-0.2$=$0.3$.
\section{Absolute Fairness and Lower Bound Results}
\label{sec:Absolute Fairness and Lower Bound Results}
\begin{figure}[h]
\centering
\includegraphics[scale=0.6]{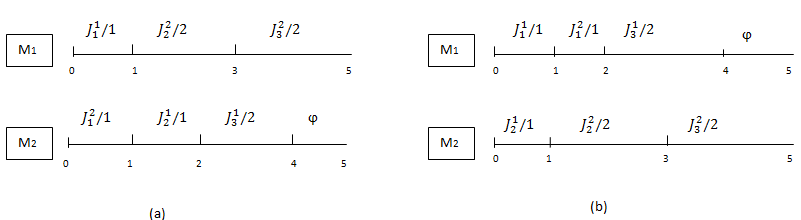}
\caption{Illustration of Case 2}
\label{fig:lowerbound.png}  
\end{figure}

We define absolute fairness as a quantitative measure and provide lower bound results for characterization of the same as follows. 
\begin{definition}
Any algorithm \textit{A} achieves \textbf{Absolute Fairness} if ${RF}(C^{r}_{A})$ is the same $\forall U_r$. 
\end{definition}
\textit{Lemma 1. If any algorithm A incurs ${RDI}^{r}_{A}$=$0$, $\forall U_r$, then it achieves absolute fairness.}\\
\textit{Proof.} If ${RDI}^{r}_{A}$=$0$, $\forall U_r$, then by Eq. (8), we have \\
\hspace*{1.4cm}${RF}(C^{r}_{A})\geq {GF}(C^{r}_{A})$\hspace*{7.1cm}(10)\\
By Eqs. (3) and (10), we can infer that ${RF}(C^{r}_{A})$=${GF}(C^{r}_{A})$, $\forall U_r$\\
Therefore Lemma 1 holds true. \hfill\(\Box\)
\begin{definition}
Any Algorithm \textit{A} is \textbf{$b$-fair}, if it achieves ${RF}(C^{r}_{A})$=$b$ for all $U_r$, where $0<b\leq 1$.
\end{definition}
\textbf{Theorem 1. Any algorithm that achieves absolute fairness for $MUMPOSP(k, P_2|C^{r}_{max})$ must be at least $(\frac{1}{k})$-fair, where $k\geq 2$ and $1\leq r\leq k$.}\\
\textit{Proof.} Let us consider an instance of $MUMPOSP(k, P_2|C^{r}_{max})$, where $k$=$2$. We analyze two cases based on $n_r$ as follows.\\
\textbf{Case $1$:} $n_1\neq n_2$.\\
Case $1$.(a): If the first job pair($J^{1}_{1}, J^{2}_{1}$) are scheduled on different machines. Let us consider the following instance $U_1:<J^{1}_2/2, J^{1}_1/1>$, $U_2:<J^{2}_{1}/1>$, where each job is specified by its processing time. Assigning $J^{1}_{1}/1$ and $J^{2}_{1}/1$ to machines $M_1$ and $M_2$ respectively followed by the assignment of $J^{1}_{2}/2$ to either of the machines such that $C^{1}_{A}$=$3$ and $C^{2}_{A}$=$1$, where $C^{1}_{OPT}\geq 1.5$ and $C^{2}_{OPT}\geq 0.5$. Therefore, we have $\frac{C^{1}_{OPT}}{C^{1}_A}\geq \frac{1}{2}$ and $\frac{C^{2}_{OPT}}{C^{2}_A}\geq \frac{1}{2}$\\
Case $1$.(b): If the first job pair($J^{1}_{1}, J^{2}_{1}$) are scheduled on the same machine. Let us consider the following instance $U_1:<J^{1}_3/2, J^{1}_2/1, J^{1}_1/1>$, $U_2:<J^{2}_{2}/2, J^{2}_{1}/1>$. If the first job pair($J^{1}_{1}/1, J^{2}_{1}/1$) are scheduled either on machine $M_1$ or on $M_2$, then by assigning the next pair of jobs ($J^{1}_{2}, J^{2}_{2}$) to the same or different machines, followed by the assignment of job $J^{1}_{3}/2$ such that $C^{1}_{A}$=$4$ and $C^{2}_{A}$=$3$, where $C^{1}_{OPT}\geq 2$ and $C^{2}_{OPT}\geq 1.5$. Therefore, we have $\frac{C^{1}_{OPT}}{C^{1}_A}\geq \frac{1}{2}$ and $\frac{C^{2}_{OPT}}{C^{2}_A}\geq \frac{1}{2}$.\\
\textbf{Case 2: $n_1$=$n_2$.}\\
Case 2.(a): If the first job pair($J^{1}_{1}, J^{2}_{1}$) are scheduled on different machines. Let us consider the following instance $U_1:<J^{1}_3/2, J^{1}_2/1, J^{1}_1/1>$, $U_2:<J^{2}_{3}/2, J^{2}_{2}/2, J^{2}_{1}/1>$. Assigning jobs $J^{1}_1/1$ and $J^{2}_{1}/1$ to machines $M_1$ and $M_2$ respectively, followed by the assignment of the subsequent jobs as shown in Figure \ref{fig:lowerbound.png}.(a), such that $C^{1}_{A}$=$4$ and $C^{2}_{A}$=$5$, where $C^{1}_{OPT}\geq 2$ and $C^{2}_{OPT}\geq 2.5$. Therefore, we have $\frac{C^{1}_{OPT}}{C^{1}_A}\geq \frac{1}{2}$ and $\frac{C^{2}_{OPT}}{C^{2}_A}\geq \frac{1}{2}$.\\
Case 2.(b): If the first job pair($J^{1}_{1}, J^{2}_{1}$) are assigned to the same machine. We consider the same instance of Case 2.(a). Assigning $J^{1}_{1}/1$ and $J^{2}_{1}$ on either machine $M_1$ or on $M_2$, followed by the assignment of the subsequent jobs as shown in Figure\ref{fig:lowerbound.png}.(b) such that $C^{1}_{A}$=$4$ and $C^{2}_{A}$=$5$. Therefore, we have $\frac{C^{1}_{OPT}}{C^{1}_A}\geq \frac{1}{2}$ and $\frac{C^{2}_{OPT}}{C^{2}_A}\geq \frac{1}{2}$. \hfill\(\Box\)
\subsection{Results on Absolute Fairness in MUMPOSP with $m$ Identical Machines for Equal Length Jobs}
For ease of understanding, we analyze the lower bound of absolute fairness for any algorithm in a generic \textit{MUMPOSP} setting, where each user has equal number of jobs and all jobs have equal processing time of $x$ unit, where $x\geq 1$. The objective of each user is to obtain a minimum $C^{r}_{max}$. We formally denote the problem as \textit{MUMPOSP($k, P_m|p^{r}_{i}$=$x|C^{r}_{max}$)}. \\\\
\textit{Lemma 2. In MUMPOSP($k, P_m|p^{r}_{i}$=$x|C^{r}_{max}$) with $k$=$b\cdot m$, any algorithm \textit{A} obtains $C^{r}_{A}\leq b\cdot \sum_{i=1}^{n_r}{p^{r}_{i}}$, for each $U_r$ respectively, where $1\leq r\leq k$, $m\geq 2$ and $b\geq 1$.}\\
\textit{Proof.} We proof Lemma 2 by method of induction on number of jobs per user ($n_r$) as follows.\\
\textit{Induction Basis:} Let us consider $k$=$m$=$2$, $n_1$=$n_2$=$1$ and $p^{1}_1$=$p^{2}_1$=$1$.
Clearly, $C^{r}_{A}$=$1 \leq b\cdot 1\cdot 1$, where $r$=$1, 2$ and $b\geq 1$.\\
\textit{Induction Hypothesis:} Let us consider $k$=$(b\cdot m)$, $n_r$=$\frac{n}{k}$=$y$, where $y\geq 1$, $b\geq 1$ and $n$=$\sum_{r=1}^{k}{n_r}$ \\
We assume that $C^{r}_{A} \leq b\cdot \sum_{i=1}^{n_r}{p^{r}_{i}}\leq b\cdot x\cdot y$ \hspace*{5.0cm} (11)
\textit{Inductive Step:} For $n_r$=$y+1$ with $p^{r}_{i}$=$x$, $\forall J^{r}_{i}$. We have to show that $C^{r}_{A} \leq (y+1)\cdot b\cdot x$. \\
By Eq. (11), we have $C^{r}_{A}$=$y\cdot b\cdot x$ with $n_r$=$y$. When we add extra one job to each user, we have by \textit{Induction Basis} $C^{r}_{A}$=$b\cdot x\cdot y+(b\cdot x)$=$(y+1)\cdot b\cdot x$. Therefore, Lemma 2 holds true. \hfill\(\Box\) \\\\
\textit{Lemma 3. Any algorithm \textit{A} is $\frac{1}{k}$-fair for MUMPOSP($P_m|p^{r}_{i}$=$x|C^{r}_{max}$) with $k$=$b\cdot m$, where $m\geq 2$ and $b\geq 1$.} \\
\textit{Proof.} By Lemma 2, we have $C^{r}_{A} \leq b\cdot \sum_{i=1}^{n_r}{p^{r}_{i}}$, $\forall U_r$ \hspace*{3.4cm}(12)\\
We have the fair optimum bound as $C^{r}_{OPT}\geq \frac{\sum_{i=1}^{n_r}{p^{r}_{i}}}{m}$, \hspace*{1.2cm}$\forall U_r$\hspace*{1.2cm} (13)\\
By Eqs. (12) and (13), we have $\frac{C^{r}_{OPT}}{C^{r}_{A}} \geq \frac{1}{k}$, \hspace*{0.1cm}$\forall U_r$.\\
Therefore, \textit{Lemma 3} holds true. \hfill\(\Box\)\\\\
\textit{Lemma 4. In MUMPOSP($P_m|p^{r}_{i}$=$x|C^{r}_{max}$) with $k> m$, any algorithm \textit{A} obtains $C^{r}_{A}\leq \lceil\frac{n}{m}\rceil \cdot x$, for each $U_r$ respectively, where $k\neq m\cdot b$ for $b\geq 1$.}\\
\textit{Proof.} The correctness of Lemma 4 is shown by method of induction on $n_r$ as follows.\\
\textit{Induction Basis:} Let us consider $m$=$2$, $k$=$3$, $n_r$=$1$ and $p^{r}_{i}$=$1$. Now, we have $n$=$n_r\cdot k$=$3$.
Clearly, $C^{r}_{A}\leq 2$=$\lceil\frac{n}{2}\rceil \cdot 1$ \\
\textit{Induction Hypothesis:} Let us consider $n_r$=$\frac{n}{k}$=$y$, $p^{r}_{i}$=$x$ and $k> m$ with $k\neq m\cdot b$ for $b\geq 1$.
We assume that $C^{r}_{A}\leq \lceil\frac{n}{m}\rceil \cdot x$, $\forall U_r$.\\
\textit{Inductive Step:} We show that $C^{r}_{A}\leq \lceil\frac{n+k}{m}\rceil \cdot x$ for $n_r$=$y+1$, $\forall U_r$.\\
By our Induction Basis, for one extra job of each user $U_r$, where $1\leq r\leq k$, algorithm \textit{A} incurs an additional time of $\lceil\frac{k}{m}\rceil \cdot x$ for each $U_r$.\\
Therefore, $C^{r}_{A}\leq \lceil\frac{n}{m}\rceil \cdot x + \lceil\frac{k}{m}\rceil \cdot x \leq \lceil\frac{n+k}{m}\rceil \cdot x$\\
Thus, Lemma 4 holds true. \hfill\(\Box\)\\\\ 
\textbf{Theorem 2. Any Algorithm \textit{A} is $\frac{1}{k}$-fair for MUMPOSP($k, P_m|p^{r}_{i}$=$x|C^{r}_{max}$), where $k\geq m$ and $m\geq 2$.}\\
\textit{Proof.} Theorem 2 holds true by Lemma 3 for $k$=$m\cdot b$, where $b\geq 1$.\\
By Lemma 4, we have $C^{r}_{A}\leq \lceil\frac{n}{m}\rceil \cdot x$ \hspace*{5.9cm} (14)\\
By Eq. (13), we have $C^{r}_{OPT}\geq \frac{\frac{n}{k}\cdot x}{m}$\\
Implies \hspace*{2.7cm} $C^{r}_{OPT}\geq \frac{n\cdot x}{k\cdot m}$ \hspace*{5.5cm}(15)\\
By Eqs. (14) and (15), we have $\frac{C^{r}_{OPT}}{C^{r}_{A}}\geq \frac{\frac{n\cdot x}{k\cdot m}}{\frac{n\cdot x}{m}}$ \\
\hspace*{5.7cm} $\geq \frac{n\cdot x\cdot m}{n\cdot k \cdot m \cdot x}\geq \frac{1}{k}$ 
\hfill\(\Box\) 
\section{Fairness Measure using Flow Time and Completion Time as User's Objective}
We show that our proposed Fairness Index can be served as a framework for measuring fairness of any algorithm based on well-known user's objectives such as \textit{sum of completion times($S^{r}$)} or weighted sum of completion times($W^{r}$) or sum of flow times($SF^{r}$). Selection of an user's objective is application dependent. For instance, users of interactive systems require optimized value for respective flow time $f^{r}$, where $f^{r}_{i}$ of any $J^{r}_{i}$ is the difference between its completion time $c^{r}_{i}$ and arrival time($t^{r}_{i}$). We now define relative fairness measures based on the above mentioned user's objectives respectively by our proposed \textit{FI}.
\begin{itemize}
\item \textbf{Sum of Completion Times($S^{r}$):} Here, the objective for each $U_r$ is to obtain a minimum $S^{r}$=$\sum_{i=1}^{n_r}{c^{r}_{i}}$. The relative fairness for any $U_r$, obtained by any algorithm \textit{A} based on $S^{r}$ is defined as\\
\hspace*{1.2cm}$R_{A}(S^{r}_{A})$=$\frac{S^{r}_{OPT}}{S^{r}_{A}}$, \hspace*{1.0cm}where $S^{r}_{OPT}$ is the optimum value for $S^{r}$.\\
\item \textbf{Weighted Sum of Completion Times($W^{r}$):} Here, the $c^{r}_{i}$ is associated with certain positive weight $w^{r}_{i}$. The objective for each $U_r$ is to obtain a minimum $W^{r}$=$\sum_{i=1}^{n_r}{w^{r}_{i}\cdot c^{r}_{i}}$. The relative fairness for any $U_r$ obtained by algorithm \textit{A} based on $W^{r}$ is defined as\\
\hspace*{1.2cm}$R_{A}(W^{r}_{A})$=$\frac{W^{r}_{OPT}}{W^{r}_{A}}$ \hspace*{1.0cm} where, $W^{r}_{OPT}$ is the optimum value for $W^{r}$.\\
\item \textbf{Sum of Flow Times(${SF}^{r}$):} Here, each $U_r$ wants a minimum value for respective ${SF}^{r}$=$\sum_{i=1}^{n_r}{f^{r}_{i}}$, where $f^{*r}_{i}$ is the desired value of $f^{r}_{i}$ and ${SF}^{r}_{OPT}$=$\sum_{i=1}^{n_r}{f^{*r}_{i}}$. The relative fairness for any $U_r$ obtained by algorithm \textit{A} based on ${SF}^{r}$ is defined as\\
\hspace*{3.2cm} $R_{A}({SF}^{r}_{A})$=$\frac{{SF}^{r}_{OPT}}{{SF}^{r}_{A}}$. 
\end{itemize} 
\section{Concluding Remarks and Scope of Future Work }
In our work, we addressed the non-trivial research challenge of defining a new fairness model with quantitative measures of algorithmic fairness for \textit{Multi-users Multi-processor Online Scheduling Problem(MUMPOSP)} based on user's objective. We formally presented the \textit{MUMPOSP} with an illustration followed by perspectives on fairness. We proposed a new fairness model and defined five quantitative measures to ensure algorithmic fairness by considering minimization of makespan as the user objective. By considering the properties of a good fairness model, our fairness measures were formally defined. We defined absolute fairness and obtained lower bound results for \textit{MUMPOSP} with identical machines for equal length jobs. We show that our proposed fairness measure can serve as a framework for measuring algorithmic fairness based on other well-known user's objective such as flow time and completion time.\\
\textbf{Scope of Future Work.} We assumed a theoretical bound for the value of $C^{r}_{OPT}$ for each $U_r$. It is still open to find a realistic bound for the value of $C^{r}_{OPT}$. It is a non-trivial research challenge to compare the fairness of any two algorithms $A$ and $B$, when global fairness of algorithm $A$ and algorithm $B$ are same and the relative fairness of $A$ is more than the relative fairness of $B$ for some users or vice versa. In this scenario, it is interesting to make a trade-off by considering the number of users and individual relative fairness of each user for comparison of fairness of two different algorithms.

%
%
%
%
%
%
%
%

\end{document}